\documentstyle{cupconf}
\input epsf

\ifoldfss
\else
  \ifnfssone
    \newmathalphabet{\mathit}
      \addtoversion{normal}{\mathit}{cmr}{m}{it}
      \addtoversion{bold}{\mathit}{cmr}{bx}{it}
    \newmathalphabet{\mathcal}
      \addtoversion{normal}{\mathcal}{cmsy}{m}{n}
    \else
    \ifnfsstwo
    \fi
  \fi
\fi

\def\etal{\mbox{\it et al.}}
\def\hexnumber#1{\ifcase#1 0\or1\or2\or3\or4\or5\or6\or7\or8\or9\or
 A\or B\or C\or D\or E\or F\fi }

\makeatletter
\ifx\CUP@mtlplain@loaded\undefined
\else
\fi
\makeatother
 \makeatletter
 \ifx\CUP@mtlplain@loaded\undefined
   \font\tenbmi=cmmib10 at 10pt
   \font\sevenbmi=cmmib10 at 7pt
   \font\fivebmi=cmmib10 at 5pt

   \newfam\bmifam
   \textfont\bmifam=\tenbmi
   \scriptfont\bmifam=\sevenbmi
   \scriptscriptfont\bmifam=\fivebmi
   
 \fi
 \makeatother
\ifnfsstwo

\fi
\ifnfssone

\fi
\ifoldfss

\fi

\mathchardef\varLambda="0103

\makeatletter
\ifx\CUP@mtlplain@loaded\undefined
\else
\fi
\makeatother
\makeatletter
\ifx\CUP@mtlplain@loaded\undefined
  \font\tenbms=cmbsy10
  \font\sevenbms=cmbsy10 at 7pt
  \font\fivebms=cmbsy10 at 5pt
  \newfam\bmsfam
  \textfont\bmsfam=\tenbms
  \scriptfont\bmsfam=\sevenbms
  \scriptscriptfont\bmsfam=\fivebms

  \edef\bsy@{\hexnumber\bmsfam}
  \mathchardef\bnabla="0\bsy@72
\fi
\makeatother
\def\etal{\mbox{\it et al.}}

\title[Evolution of bulge shapes]{Dynamical evolution of bulge shapes}

\author[M. Valluri]%
{M\ls O\ls N\ls I\ls C\ls A\ns     V\ls A\ls L\ls L\ls U\ls R\ls I}
\affiliation{Department of Physics and Astronomy, Rutgers University,
136 Frelinghuysen Road, Piscataway, NJ 08854-8019}
\begin{document}
\ifnfssone
\else
  \ifnfsstwo
  \else
    \ifoldfss
      \let\mathcal\cal
      \let\mathrm\rm
      \let\mathsf\sf
    \fi
  \fi
\fi

\maketitle

\begin{abstract}
Figure rotation substantially increases the fraction of stochastic
orbits in triaxial systems. This increase is most dramatic in systems
with shallow cusps showing that it is not a direct consequence of
scattering by a central density cusp or black hole.  In a recent study
of stationary triaxial potentials (\cite{val98}) it was found that the
most important elements that define the structure of phase space are
the two dimensional resonant tori.  The increase in the fraction of
stochastic orbits in models with figure rotation is a direct
consequence of the destabilization of these resonant tori.
 
The presence of a large fraction of stochastic orbits in a triaxial
bulge will result in the evolution of its shape from triaxial to
axisymmetric. The timescales for evolution can be as short as a few
crossing times in the bulges of galaxies and evolution is accelerated
by figure rotation.  This suggests that low luminosity ellipticals and
the bulges of early type spirals are likely to be predominantly
axisymmetric.

\end{abstract}

\firstsection 
\section{Introduction}

It is now widely believed that the effects of central black holes and
cusps on the dynamics of triaxial galaxies are well understood: the
box orbits which form the back bone of triaxial elliptical galaxies
become chaotic due to scattering by the divergent central force
(e.g. \cite{ger85}).  The scattering of these orbits then results in
the evolution of the triaxial galaxy to an axisymmetric one whose
dynamics is dominated by well behaved families of regular orbits. Thus
most studies of elliptical galaxies still focus on the nature of the
regular orbits. Recent investigations of the structure of phase space
in triaxial ellipticals have shown that phase space is rich in regular
and chaotic regions even in the absence of black holes and steep cusps.

Studying the effects of central black holes on galaxies has taken on
renewed importance because of the discovery that many if not most
bulge dominated galaxies have central black holes.
 The existence of central black holes as the end products of the QSO
and AGN phenomena is justified by energetic arguments. But less is
known about the interplay between the growth of a black hole and the
shape of its host galaxy. Most models for the fueling of QSO and AGN
require a high degree of triaxiality to transport fuel to the center
and to simultaneously transport angular momentum outwards
(\cite{ree90}).  Understanding the interplay between black hole growth
and galaxy shape is one motivation for studying the behavior of orbits
in triaxial potentials.

There have been several studies of the effect of figure rotation on
the orbits of stars in triaxial galaxies. Most studies have focused on
the behavior of the periodic orbits in the plane perpendicular to the
rotation axis.  Some authors (\cite{mar90}) found that increasing
figure rotation resulted in a decrease in the phase space occupied by
the unstable $x_3$ family and consequently a reduction in the overall
chaos.  Others (\cite{udr88} and \cite{udr91}) found that increasing
figure rotation had negligible effect on the stochasticity of orbits
in 3-dimensional models. More recently it has been shown
(\cite{tsu93}) that orbits of all 4 major families in a perfect
ellipsoidal model (completely integrable when stationary) became
stochastic when figure rotation is added.  Rapidly rotating triaxial
bars can be almost completely regular (Pfenniger \& Friedli 1991)
although more slowly rotating bars and bars with high central
concentrations generally contain a large fraction of stochastic orbits
that eventually destroy the bars (\cite{nor96},
\cite{sel99}).

We use the frequency analysis technique (\cite{las90}) to study the
behavior of orbits in a family of triaxial density models with figure
rotation. The models have a density law that fits the observed
luminosity profiles of ellipticals and the bulges of spirals and is
given by Dehnen's law
\begin{equation}
\rho(m) = {(3-\gamma) M\over 4\pi abc} m^{-\gamma} 
(1+m)^{-(4-\gamma)},
\ \ \ \ 0 \le \gamma < 3
\label{deh1}
\end{equation}
where
\begin{equation}
m^2={x^2\over a^2} + {y^2\over b^2} + {z^2\over c^2}, \ \ \ \ 
a\ge b\ge c\ge
0
\label{deh2}
\end{equation}
and $M=1$ is the total mass.  The parameter $\gamma$ determines the
slope of the central density cusp and $a, b, c$ are the semi-axes of
the model.  In some cases we also introduced a central point mass
$M_h$ representing a nuclear black hole.  The figure rotates about its
short axis and the degree of figure rotation can be small (as in the
case of giant ellipticals) or reasonably large as in the case of
bulges. The co-rotation radius $R_\Omega$ is parameterized in units of
the half-mass radius of the model and ranges from $R_\Omega = 25$
(slowly rotating) to $R_\Omega = 3$ (rapidly rotating). Frequency
analysis was restricted to $\sim 10^4$ orbits in each model. Orbits
were launched from the equi-effective-potential surface corresponding
to the half-mass radius.  (Thus all orbits have the same Jacobi
Integral, $E_J = E - {1\over{2}}|{\bf\Omega\times r}|^2$).  The
initial conditions for the orbits were selected in two different
ways to study  orbits from all four major families.

\section{Frequency mapping and resonant tori}

Laskar's (1990) frequency analysis technique is based on the idea that
regular orbits have 3 isolating integrals of motion which are related
to 3 fundamental frequencies.  A filtered Fourier transform technique
can be used to accurately determine these 3 frequencies ($\omega_x$,
$\omega_y$, $\omega_z$). While stochastic orbits do not really have
fixed frequencies, quantities resembling frequencies which measure
their local behavior can be used to determine how they diffuse in
frequency space.  Regular orbits come in three types: (1) Orbits in
regions that maintain their regular character in spite of departures
of the potential from integrable form; (2) orbits associated with
stable resonant tori; (3) orbits associated with stable periodic
orbits, or ``boxlets''.

The use of frequency mapping has shown that even in weakly chaotic
systems, it is the {\it resonant tori} that provide the skeletal
structure to regular phase space (\cite{val98}). 
Frequency mapping provides the simplest method for finding resonant
tori.  They are families of orbits which satisfy a condition:
$l\omega_x + m\omega_y + n\omega_z = 0$ with $(l,m,n)$ integers. Such
orbits are restricted to 2-dimensional surfaces in phase space and we
refer to them as {\it thin orbits}. Thin boxes are the most generic
box orbits in non-integrable triaxial potentials. They avoid the
center because they are two-dimensional surfaces. They generate
families of 3-D boxes whose maximum thickness is determined by the
strength of the central cusp or black hole (\cite{mer99}). The closed
periodic boxlet orbits lie at the intersection of two or more
resonance zones. High order resonances also exist for tube orbit
families. Unlike the well known thin tube families around the long and
short axes, thin resonant tubes are often surrounded by unstable
regions, making it difficult to find them without a technique like
frequency mapping.

%
%
\begin{figure}
  \epsfysize=3.in
  \centerline{  \epsfbox{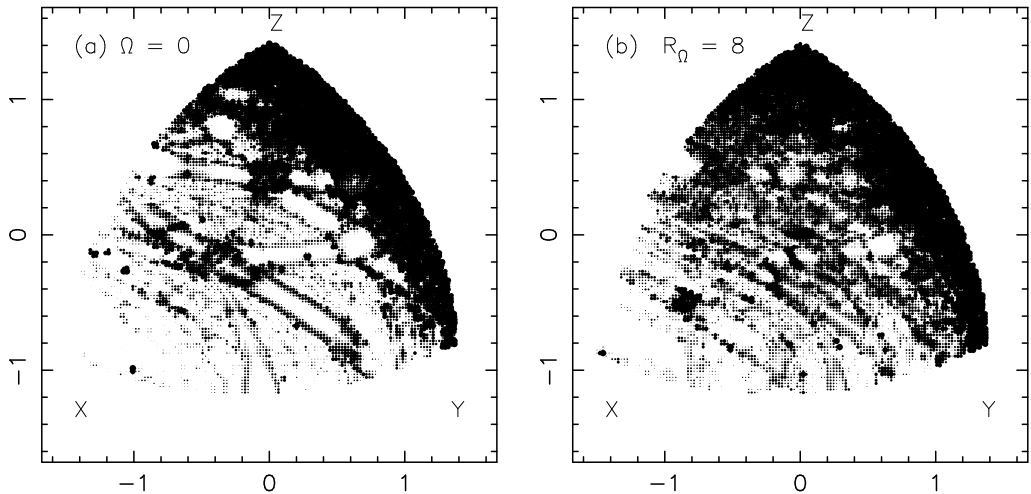}}
  \caption{(a) The initial-condition-space diffusion map of box-like
orbits in the non-rotating model.
(b) Like (a) for orbits started on the
equi-effective-potential surface of a slowly rotating model with
$R_{\Omega} = 8$. Features on the maps are described in the text}
\label{fig:diffmap}
\end{figure}

\section{Results: Destruction of the Resonant Tori}

A box or boxlet orbit reverses its sense of progression around the
rotation axis every time it reaches a turning point. In a rotating
frame this means that the path described during the prograde segment
of the orbit is not retraced during the retrograde segment. This
``envelope doubling'' is a consequence of the Coriolis forces on the
two segments being different (\cite{dez83}).  Envelope doubling
effectively thickens the thin box orbits driving them closer to the
center. This results in a narrowing of the stable portion of the
resonance layer and renders a large fraction of the orbits stochastic.
The degree of ``thickening'' increases with increasing figure rotation
and results in a corresponding rise in the fraction of  stochastic
box-like orbits.

 Figure~1 (a) shows a plot of a quantity measuring the diffusion rates
of $10^4$ orbits started at rest at the half-mass equi-potential
surface in a non-rotating triaxial model with central cusp slope
$\gamma = 0.5$. Only one octant of the surface is plotted. The grey
scale is proportional to the logarithm of the diffusion rate: the dark
regions indicate initial conditions corresponding to stochastic
orbits, the white regions correspond to regular orbits.  Figure~1 (b)
shows the same set of orbits started from the equi-effective-potential
surface of a model with $R_\Omega = 8$. Rotation results in the
broadening of the unstable regions with a resultant narrowing of the
stable (white) regions. It also gives rise to new unstable and stable
resonances which are seen in Figure~1 (b) as dark striations within
the white regions. The increase in the number of resonances and their
broadening results in greater overlap of nearby stochastic layers
eventually leading to the onset of global stochasticity
(e.g. \cite{chi79}).
 
Contrary to the finding of Tsuchiya \etal~\ (1993) we find that figure
rotation has a strong destabilizing effect on inner-long axis
tubes. 
The low angular momentum $z$-tubes and the outer $x$-tubes also become
more stochastic. The high angular momentum $z$-tubes are much less
affected. The increased stochasticity of tube orbits can be attributed
largely to the increase in the width of the stochastic layers
associated with the resonant tube orbit families. We emphasize that
for the tube orbits it is the destabilization of resonant tubes and
not scattering by divergent central forces that determines their
stability.

\section{Conclusions} 

It is a popular misconception that in the presence of figure rotation
box orbits in a triaxial elliptical will loop around the center due
to Coriolis forces thereby reducing stochasticity.  We find that on the
contrary stochasticity increases with increasing figure rotation
primarily because the thin box orbits and resonant tubes, which play a
crucial role in structuring phase space, are broadened and
destabilized by the ``envelope doubling'' effect.

Models for the fueling of AGN and QSOs require triaxial central
potentials which aid accretion onto a black hole, but the same black
holes would tend to destroy triaxiality. Low luminosity ($M_B > -19$)
ellipticals and the bulges of spirals are expected to evolve into
axisymmetric shapes on time scales much shorter than the age of the
Universe (\cite{val98}).  If the peanut-shaped bulges in nearby
galaxies are in fact triaxial they are probably dynamically young or are
composed of only tube like orbits.

\begin{acknowledgments}
I thank David Merritt for useful discussions.  This work was supported
 by NSF grants AST 93-18617 and AST 96-17088 and NASA grant NAG 5-2803
 to Rutgers University.

\end{acknowledgments}

\end{document}